\documentclass[prb,superscriptaddress,showpacs,12pt]{revtex4-1}

\usepackage{amssymb, amsmath}

\usepackage{float}

\usepackage{graphicx}
\usepackage{color}

\begin{document}

\title{Investigation of the physical properties of the Ce$_{2}$MAl$_{7}$Ge$_{4}$ (M = Co, Ir, Ni, Pd) heavy fermion compounds }

\author{N. J. Ghimire}
\email{nghimire@lanl.gov}
\affiliation{Los Alamos National Laboratory, Los Alamos, NM 87545, USA}

\author{S. K. Cary}
\affiliation{Department of Chemistry and Biochemistry, Florida State University, Tallahassee, FL 32306, USA}

\author{S. Eley}
\affiliation{Los Alamos National Laboratory, Los Alamos, NM 87545, USA}

\author{N. A. Wakeham}
\affiliation{Los Alamos National Laboratory, Los Alamos, NM 87545, USA}

\author{P. F. S. Rosa}
\affiliation{Los Alamos National Laboratory, Los Alamos, NM 87545, USA}

\author{T. Albrecht-Schmitt}
\affiliation{Department of Chemistry and Biochemistry, Florida State University, Tallahassee, FL 32306, USA}

\author{Y. Lee}
\affiliation{Center for Artificial Low Dimensional Electronic Systems, Institute for Basic Science, Department of Physics at POSTECH Pohang, Korea 790-784}

\author{M. Janoschek}
\affiliation{Los Alamos National Laboratory, Los Alamos, NM 87545, USA}

\author{C. M. Brown}
\affiliation{National Institute of Standards and Technology, Center for Neutron Research, Gaithersburg, MD 20878 USA}

\author{L. Civale}
\affiliation{Los Alamos National Laboratory, Los Alamos, NM 87545, USA}

\author{J. D. Thompson}
\affiliation{Los Alamos National Laboratory, Los Alamos, NM 87545, USA}

\author{F. Ronning}
\affiliation{Los Alamos National Laboratory, Los Alamos, NM 87545, USA}

\author{E. D. Bauer}
\affiliation{Los Alamos National Laboratory, Los Alamos, NM 87545, USA}

\date{\today}

\begin{abstract}
 We report the synthesis, crystal structure and characterization by means of single crystal x-ray diffraction, neutron powder diffraction, magnetic, thermal and transport measurements of the new heavy fermion compounds Ce$_{2}$MAl$_{7}$Ge$_{4}$ (M = Co, Ir, Ni, Pd). These compounds crystallize in a noncentrosymmetic tetragonal space group    P\={4}2$_{1}$m, consisting of layers of square nets of Ce atoms separated by Ge-Al and M-Al-Ge blocks. Ce$_{2}$CoAl$_{7}$Ge$_{4}$, Ce$_{2}$IrAl$_{7}$Ge$_{4}$ and Ce$_{2}$NiAl$_{7}$Ge$_{4}$ order magnetically behavior below $T_{M}=$ 1.8, 1.6, and 0.8 K, respectively.  There is no  evidence of magnetic ordering in Ce$_{2}$PdAl$_{7}$Ge$_{4}$ down to 0.4 K. The small amount of entropy released in the magnetic state of Ce$_{2}$MAl$_{7}$Ge$_{4}$ (M = Co, Ir, Ni) and the reduced specific heat jump at $T_M$ suggest a strong Kondo interaction in these materials. Ce$_{2}$PdAl$_{7}$Ge$_{4}$ shows non-Fermi liquid behavior, possibly due to the presence of a nearby quantum critical point.
\end{abstract}
\maketitle

\section{Introduction}

In heavy fermion materials, hybridization between f-electron and conduction electron states gives rise to a large specific heat coefficient $C/T=\gamma$, often 100-1000 times that of ordinary metals (e.g., Cu).\cite{Stewart1984,Fisk1988}  The ground state is determined by a balance of competing Kondo and RKKY interactions, which are both described by an exchange interaction $|\mathcal{J}|$ that characterizes the hybridization strength between f-electron/conduction-electron states, as proposed by Doniach.\cite{Doniach77} When $|\mathcal{J}|$ is small, the RKKY interaction dominates, and the system orders magnetically, but for large $|\mathcal{J}|$, the material behaves as a heavy Fermi liquid: $\gamma \sim $ const., magnetic susceptibility $\chi \sim $ const. and electrical resistivity $\rho \sim T^2$.  When the RKKY and Kondo interactions become comparable, the magnetic phase transition is suppressed to zero temperature at a quantum critical point (QCP).\cite{vonLohneysen07,Sachdev2011,Stockert2011} Here, quantum fluctuations give rise to  non-Fermi liquid (NFL) behavior in the physical properties, e.g.,  $C/T \sim -ln(T)$ or $\sim T^{-n}$, $\chi \sim \ln(T)$ or $\sim T^{-n}$, and  $\rho \sim T^n$ ($n<2$), depending on the nature of the fluctuations. A dome of unconventional superconductivity often occurs in the vicinity of the quantum critical point, where the quantum fluctuations are strongest and presumably mediate the superconductivity.\cite{Monthoux07,vonLohneysen07}  To explore the interplay of superconductivity and magnetism near a QCP, attention has focused on several families of tetragonal Ce-based compounds. For example, superconductivity  is observed near an antiferromagnetic quantum critical point in CeM$_2$X$_2$ (M = transition metal; X=Si, Ge),\cite{Pfleiderer2009}  the Ce$_{m}$M$_{n}$In$_{3m+2n}$ family,\cite{Thompson2003} (M = transition metal, \emph{m} and \emph{n} are the number of the CeIn$_{3}$  and MIn$_{2}$ building blocks, respectively), and  non-centrosymmetric CePt$_3$Si and CeMX$_{3}$ (M= transition metal; X=Si, Ge) compounds.\cite{Bauer2012book}

The local crystal-chemical environment  influences the magnetic behavior of  heavy fermion materials.\cite{Szytula1994} When a rare earth ion, like Ce, is placed within a particular crystal-chemical environment, the interactions with the surrounding ligands (e.g., crystalline electric field (CEF), hybridization) modify the free-ion 4f electron wavefunction  and its electronic and magnetic properties. Conventional density functional theory (DFT) calculations, unfortunately, are unable to capture sufficiently the complexity of these interactions with the detail that is needed to predict the outcome of these interactions.\cite{Lejaeghere2016,Skylaris2016} To our knowledge, theory has not anticipated any material to exhibit heavy fermion properties.  Instead, progress in understanding the interplay between crystal structure and chemistry and in the discovery of interesting new heavy fermion states has come from an iterative process of materials discovery, characterization and theory/modeling. The materials mentioned above are good examples of this paradigm and have been particularly instructive because they each comprise a family in which interactions can be tuned systematically by the crystal-chemical environment of the Ce's $4f$ electron.  Only after these materials were discovered did experimental and theoretical study show how interesting they were.  A crystal-chemical lesson from them is that a fruitful direction to look for similarly interesting behaviors is to explore for new Ce-containing materials with symmetry lower than cubic and containing a transition metal as well as an element from IIIB or IVB columns in the periodic table.  This lesson is born out in a more recent example of the quasi-two dimensional CeMAl$_4$Si$_2$ family (M=Rh, Ir, Pt) in which nominally isoelectronic Rh and Ir members are moderately heavy fermion antiferromagnets and nominally electron-doped M=Pt is ferromagnetic.\cite{Ghimire2015a,Ghimire2015b,Maurya2016}  In the course of exploring the possible consequences of expanding the crystal lattice of this family by replacing Si with larger Ge, we discovered a new family of quasi-two dimensional Ce$_{2}$MAl$_{7}$Ge$_{4}$ (M=Co, Ir, Ni. Pd) compounds that, as we discuss, are interesting in their own right.     Herein, we report the synthesis and physical characterization of single crystals of  this new family of compounds by means of x-ray diffraction, magnetization, electrical resistivity, and specific heat.  Unlike the CeMAl$_4$Si$_2$ materials, these Ce$_{2}$MAl$_{7}$Ge$_{4}$ compounds crystallize in a noncentrosymmetic tetragonal space group    P\={4}2$_{1}$m.  This crystal-chemical environment  leads to  magnetic order below $T_{M}$=1.8, 1.6, and 0.8 K, in Ce$_{2}$CoAl$_{7}$Ge$_{4}$, Ce$_{2}$IrAl$_{7}$Ge$_{4}$ and Ce$_{2}$NiAl$_{7}$Ge$_{4}$, respectively.   A large $4f$ contribution to the specific heat  and the reduced entropy below $T_{M}$ [$\sim$ 0.2-0.3 Rln(2), where R is the gas constant] suggest a significant Kondo interaction in these three  materials;  the comparable Kondo and RKKY energy scales  indicate that the Ce$_{2}$MAl$_{7}$Ge$_{4}$ compounds are close to the magnetic/nonmagnetic boundary in the Doniach diagram.\cite{Doniach77}  Ce$_{2}$PdAl$_{7}$Ge$_{4}$ does not order magnetically  and shows non-Fermi liquid behavior down to 0.4 K with the Sommerfeld coefficient reaching $\gamma \sim 1$ J/mol-Ce K$^2$. There is no detectable superconductivity in the Ce$_{2}$MAl$_{7}$Ge$_{4}$ materials above 0.4 K.

\section{Experimental details}

Single crystals of Ce$_{2}$MAl$_{7}$Ge$_{4}$ (M = Co, Ir, Ni, Pd) were grown from an Al/Ge flux. First, CeMGe$_{3}$ was prepared by arc-melting the stoichiometric composition on a water-cooled copper hearth under an argon atmosphere with a Zr getter. The arc-melted  CeMGe$_{3}$ was mixed with Al$_{88}$Ge$_{12}$ in a ratio of 1:8 by weight and loaded into a 5-ml crucible. The crucible was sealed in a  quartz ampoule under vacuum. The sealed ampoule was heated to 1100$^{\circ}$C in 6 hours, homogenized at 1100$^{\circ}$C for 24 hours and then slowly cooled to 700$^{\circ}$C at the rate of 4$^{\circ}$C/hour. Once the furnace reached 700$^{\circ}$C, the excess flux was decanted from the crystals using a centrifuge. Plate-like crystals were obtained in all cases. Crystals as large as 8$\times$6$\times$0.3 mm were obtained (Fig. \ref{Figure1}d). Isostructural La$_{2}$MAl$_{7}$Ge$_{4}$  analogs  were  prepared by a similar method.

The crystal structure of the Ce$_{2}$MAl$_{7}$Ge$_{4}$ compounds was determined by single crystal x-ray diffraction at room temperature. Large single crystals were isolated and broken into small fragments, which were then mounted on Mitogen loops with Immersion oil, and optically aligned on a Bruker D8 Quest X-ray diffractometer using a digital camera.  Initial intensity measurements were performed using an I$\mu$S X-ray source (MoK$_{\alpha}$, $\lambda$ = 0.71073 \AA) with high-brilliance and high-performance focusing multilayered optics. Standard APEXII software was used for determination of the unit cells and data collection control. The intensities of reflections of a sphere were collected by a combination of multiple sets of exposures (frames).  Each set had a different $\varphi$ angle for the crystal and each exposure covered a range of 0.5$^{\circ}$ in $\omega$. A total of 1464 frames were collected with an exposure time per frame of 5 to 10 seconds, depending on the sample. SAINT software was used for data integration including Lorentz and polarization corrections. The large size of the crystals necessitated breaking the crystals and using small fragments as described above, which precludes a face-indexed numerical absorption correction.

Neutron powder diffraction experiments were carried out on  Ce$_{2}$CoAl$_{7}$Ge$_{4}$ and  Ce$_{2}$NiAl$_{7}$Ge$_{4}$ using the high resolution powder diffractometer BT-1 at the NIST Center for Neutron Research.\cite{NIST} A neutron wavelength of 2.0775 \AA{} was selected using a Ge(311) monochromator, which avoids  higher-order contamination of the monochromated beam via its crystal symmetry. Powder samples were prepared by pulverizing single crystals. 2.4 grams of Ce$_{2}$CoAl$_{7}$Ge$_{4}$ and 3.0 grams of Ce$_{2}$NiAl$_{7}$Ge$_{4}$ were loaded in aluminium and copper cans, respectively. A diffraction pattern was collected for 20 hours  at 4.2 and 0.42 K for Ce$_{2}$CoAl$_{7}$Ge$_{4}$ and at 2, 0.5 and 0.137 K for Ce$_{2}$NiAl$_{7}$Ge$_{4}$. A Rietveld refinement of the crystal structure was carried out using the program FullProf. \cite{Fullprof}

The chemical composition of the Ce$_{2}$MAl$_{7}$Ge$_{4}$ single crystals was analyzed using a scanning electron microscope (SEM) with an energy dispersive X-ray spectrometer (EDS). The atomic percentages of the Ce$_{2}$MAl$_{7}$Ge$_{4}$ compounds were determined to be: Ce, Co, Al, Ge=14.22, 6.82, 49.93, 29.03; Ce, Ir, Al, Ge=14.46, 7.31, 47.93, 30.30; Ce, Ni, Al, Ge=14.60, 7.42, 48.13, 29.80; and  Ce, Pd, Al, Ge=14.49, 7.30, 47.71, 30.00.  These values are within the expected uncertainty for the standardless measurements, fully consistent with the composition Ce:M:Al:Ge = 2:1:7:4, normalized to M, and with the site occupancies found in refinement of the X-ray diffraction measurements in the P\={4}2$_{1}$m space group (see Appendix). The La-analogs were also characterized by chemical composition analysis with similar results.
DC magnetization measurements between 0.4 K and 350 K in magnetic fields up to H=70 kOe were performed in a Quantum Design Magnetic Property Measurement System (MPMS).  Electrical resistivity and specific heat capacity measurements of crystals were carried out in a Quantum Design Physical Property Measurement System (PPMS) down to 0.4 K and in magnetic fields up to 90 kOe. The heat capacity of the La-compounds was measured down to 2 K. Electrical contacts for resistivity measurements were made by spot welding 25-$\mu$m diameter Pt wires onto the sample so that resistivity was measured with current applied in the ab-plane. The electrical resistivity of Ce$_{2}$CoAl$_{7}$Ge$_{4}$ and  Ce$_{2}$NiAl$_{7}$Ge$_{4}$ was measured in an external field of 200 and 500 Oe, respectively, to suppress the superconductivity due to traces of Al-Ge impurities.\cite{Roberts1976}

\section{Results}

\subsection{Crystal Structure}
\begin{figure}[H]
\begin{center}
\includegraphics[scale=.8]{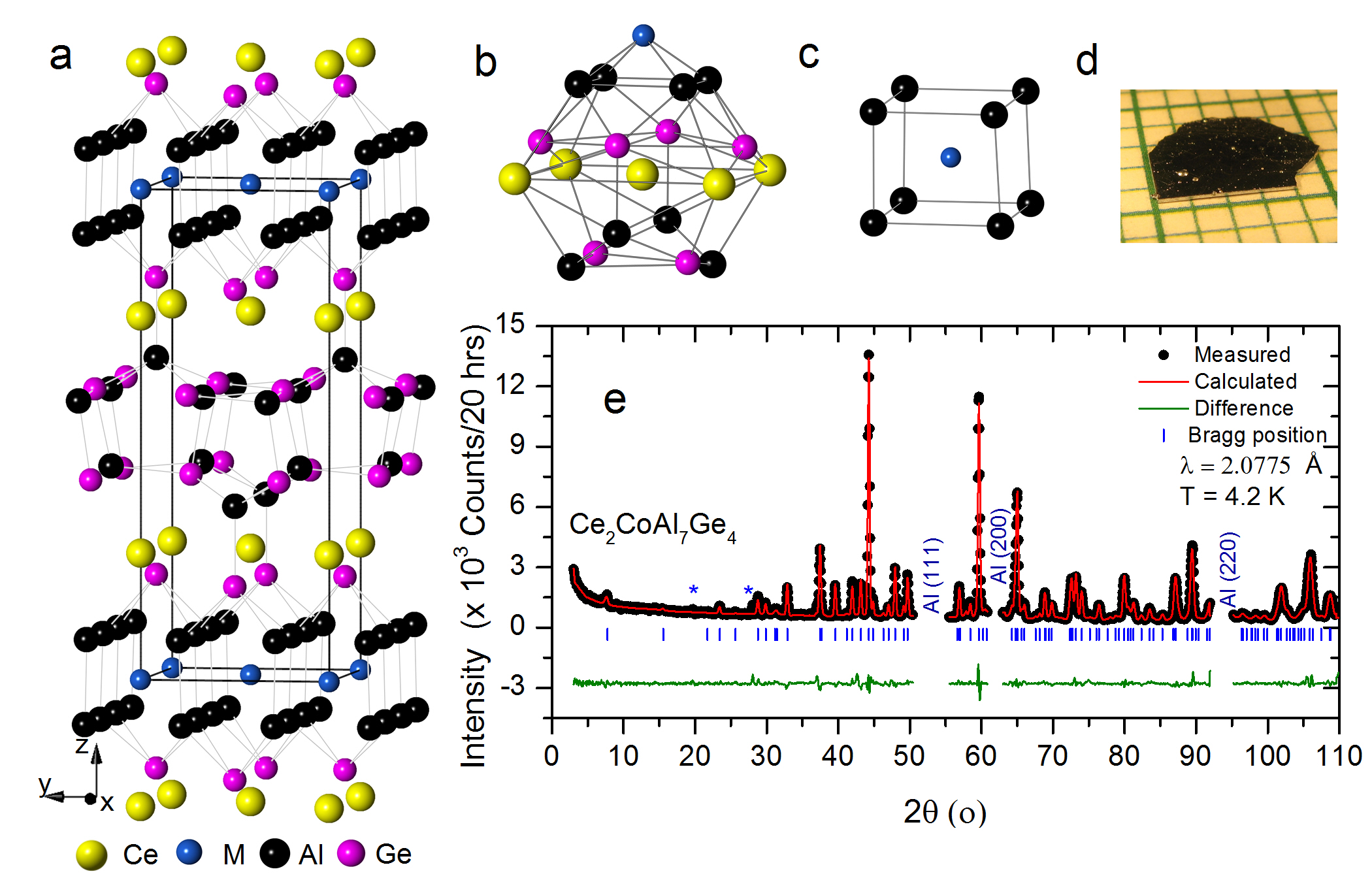}
\caption{(Color online) (a) Crystal structure of Ce$_{2}$MAl$_{7}$Ge$_{4}$, b) Atomic environment of Ce, c) Atomic environment of $M$, d) Photo of a crystal of Ce$_{2}$NiAl$_{7}$Ge$_{4}$ on a 2 mm $\times$ 2 mm grid, and e) Rietveld refinement of the neutron powder pattern of Ce$_{2}$CoAl$_{7}$Ge$_{4}$ measured above the magnetic transition temperature at T=4.2 K. The Al peaks from the sample holder are excluded. The asterisks ($\ast$)  indicate unknown impurity peaks.}\label{Figure1}
\end{center}
\end{figure}

\begin{table}[h]

\caption{Crystallographic data of Ce$_{2}$MAl$_{7}$Ge$_{4}$ (M = Co, Ir, Ni, Pd) determined by single crystal x-ray diffraction. }\label{Table-crystal1}
\begin{tabular*}{1\linewidth}{@{\extracolsep{\fill}}ll}
\hline
   Crystal system            &                Tetragonal	$\hspace{10.0cm}$                                                                                                    \\
   Space group               &         	   P\={4}2$_{1}$m	                                                                                                \\
   Temperature               &                296(2) K                                                                                                                            \\
   Wavelength                &                0.71073 \AA                                                                                                       \\
   Z formula units           &                2                                                                                                               \\
   2$\theta_{min}$           &                2.65$^{o}$                                                                                                         \\
   2$\theta_{max}$           &                27.55$^{o}$                                                                                                       \\
   Index ranges              &                -7 $\leqslant$ h $\leqslant$ 7,    -7 $\leqslant$ k $\leqslant$ 7, -19 $\leqslant$ l $\leqslant$ 20              \\
\end{tabular*}
\begin{tabular*}{1\linewidth}{@{\extracolsep{\fill}}lcccc}
\hline
Compound	 &   Ce$_{2}$CoAl$_{7}$Ge$_{4}$	  &	   Ce$_{2}$IrAl$_{7}$Ge$_{4}$	&	Ce$_{2}$NiAl$_{7}$Ge$_{4}$ & Ce$_{2}$PdAl$_{7}$Ge$_{4}$\\
\hline							
Formula weight                          &	818.39 	         &	  951.66 	    &	818.17            & 865.86	       \\
  a (\AA)\                               &	5.9158(3) 	     &	  5.9786(4) 	&	5.9343(3)         & 6.0070(4)	       \\
  c (\AA)\	                            &	15.3635(19) 	 &	  15.366(2) 	&	15.2733(17)       & 15.2632(19)	        \\
  Volume (\AA$^{3}$)	                &	537.67(8) 	     &	  549.25(9)  	&	537.86(7)         & 550.76(9)      \\
  Density (calculated)(g/cm$^{3}$)      &	5.055  	         &	  5.754   	    &	5.052             & 5.221          \\
  $\mu$ (Mo K$\alpha$) (cm$^{-1}$)      &    213.26           &    314.69       &   215.27            & 209.47         \\
  Goodness-of-fit on F$^{2}$            &	1.127  	         &	  1.128   	    &	1.245             & 1.211          \\
  $R(F)$ for $F_{o}^{2}$ $>$            &                   &                   &                    &                       \\
  2$\sigma (F_{o}^{2})^{a}$             &    0.0171          &    0.0175        &   0.0186            & 0.0265          \\
  $R_{w}(F_{o}^{2})^{b}$                &   0.0422  	         &	  0.0401   	    &	0.0459            & 0.0700          \\
\hline
\hline
\end{tabular*}
\begin{tabular*}{1\linewidth}{@{\extracolsep{\fill}}ll}
 $^{a}R(F)$ = $\sum\mid\mid F_{o}\mid -\mid F_{c}\mid\mid/\sum\mid F{_o}\mid$ &  $^{b}R_{w}(F_{o}^{2})$	= $[\sum[w(F_{o}^{2}-F_{c}^{2})^{2}]/\sum wF_{o}^{4}]^{1/2}$   \\ 	 \end{tabular*}				
\end{table}

\begin{table}[h]
\caption{Atomic coordinates and equivalent displacement parameters (\AA$^{2}$) for Ce$_{2}$CoAl$_{7}$Ge$_{4}$ determined by single crystal x-ray diffraction at room temperature. U$_{eq}$ is defined as one third of the trace of the orthogonalized U$^{ij}$ tensor.}\label{Table-Co}
\begin{center}
\begin{tabular}{l@{\hspace{0.7cm}}c@{\hspace{0.7cm}}c@{\hspace{0.7cm}}c@{\hspace{0.7cm}}c@{\hspace{0.7cm}}c@{\hspace{0.7cm}}c}								
		\hline													
        Atom           & Wyck.      & Occ.   &	  x	          &	      y	         &	  z	        &   U$_{eq}$    \\
		\hline													
         Co(1)	       &	2a	    &	1	&	0	          &	0	             &	0	        &	0.005(1)	\\
         Ce(1)	       &	4d	    &	1	&	0	          &	0	             &	-0.2564(1)	&	0.006(1)	\\
         Ge(1)	       &	2c	    &	1	&	0.5	          &	0	             &	-0.2008(1)	&	0.008(1)	\\
         Ge(2)	       &	2c	    &	1	&	0	          &	0.5	             &	-0.1892(1)	&	0.008(1)	\\
         Ge(3)	       &	4e	    &	1	&	0.2810(1)	  &	-0.2190(1)	     &	-0.4124(1)	&	0.011(1)	\\
         Al(1)	       &	4e	    &	1	&	-0.2920(3)	  &	-0.2080(3)	     &	-0.4279(1)	&	0.007(1)	\\
         Al(2)	       &	2c	    &	1	&	0	          &	-0.5	         &	-0.3531(2)	&	0.009(1)	\\
         Al(3)	       &	4e	    &	1	&	-0.2557(2)	  &	-0.2443(2)	     &	-0.0835(2)	&	0.008(1)	\\
         Al(4)	       &	4e	    &	1	&	-0.7436(2)	  &	-0.2436(2)	     &	-0.0841(2)	&	0.010(1)	\\
 					
	  \hline      					
	    \hline
\end{tabular}
  \end{center}
\end{table}

Analysis of single crystal X-ray diffraction patterns revealed that the compounds Ce$_{2}$MAl$_{7}$Ge$_{4}$ (M = Co, Ir, Ni, Pd) are isostructural and crystallize in a defect-variant of the Sm$_2$NiGa$_{12}$ structure-type\cite{Chen2000,Cho2008,Macaluso2005} with a noncentrosymmetric space group P\={4}2$_{1}$m (\#113). 
Any possible missed symmetry in the structure of the Ce$_{2}$MAl$_{7}$Ge$_{4}$ compounds was checked for using the Platon software package\cite{Platon} along with CHECKCIF, both of which indicated possible pseudosymmetry and a possible centrosymmetric space group of P4/nmm. However, the reduction of symmetry to the noncentrosymmetric, but not polar, P\={4}2$_{1}$m space group is invoked by the differences in the Ce-Al vs Ce-Ge bond distances, which approximates, but does not equal, an n-glide.  Furthermore, the refinement in P4/nmm is considerably poorer than in P\={4}2$_{1}$m and also induces Al/Ge disorder that, upon refinement, leads to a formula inconsistent with the microprobe analysis described below.  In addition, the thermal displacement parameters are unreasonable and in some cases nearly non-positive definite in P4/nmm.  All indications are that P\={4}2$_{1}$m is the correct space group. Crystallographic Information Files (CIFs) are available from the Inorganic Crystal Structure Database (ICSD).\cite{ICSD}

The crystallographic data are presented in Table \ref{Table-crystal1} and the fractional atomic coordinates and equivalent displacement parameters for Ce$_{2}$CoAl$_{7}$Ge$_{4}$ are given in Table \ref{Table-Co} (the atomic coordinates of the M=Ni, Pd, and Ir compounds are listed in the Appendix in Tables \ref{Table-Ir}-\ref{Table-Pd}). A unit cell of Ce$_{2}$MAl$_{7}$Ge$_{4}$ consists of 4 Ce atoms, 2 M atoms, 14 Al atoms and 8 Ge atoms (two formula units per unit cell, \emph{Z}=2). The crystallographically equivalent Ce atoms occupy the $4d$  Wyckoff position. The M atoms are also crystallographically equivalent residing in $2a$ Wyckoff position. There are three crystallographically inequivalent Ge sites and four crystallographically inequivalent Al sites where Ge1, Ge2, Ge3 occupy the $2c$, $2c$ and $4e$ Wyckoff positions and Al1, Al2, Al3 and Al4 occupy the $4e$, $2c$, $4e$ and $4e$ Wyckoff positions, respectively. The atomic environment  of Ce and M atoms constructed with atoms before the maximum gap \cite{Daams1997} are depicted in Figs. \ref{Figure1}b and c. Each Ce atom is surrounded by eight Al, six Ge, four Ce and one M atoms with a coordination number of 19. Each M atom is surrounded by eight Al atoms in a nearly cubic environment. Thus, one may view the crystal structure as consisting of a planar square net of Ce atoms separated  by M-Al-Ge and Al-Ge blocks alternating along the c-axis as shown in Fig. \ref{Figure1}a. The Ce-Ge and Ce-Al distances range from 3.1016(3) to 3.1640(6) \AA{} and 3.3071(10) to 3.3816(19) \AA, respectively.

The replacement of some Ge with Al within the local environment around Ce sites causes a reduction in the site symmetry from 2/m in  Sm$_2$NiGa$_{12}$  to simple twofold symmetry.  However, the M sites are surrounded solely by Al in Ce$_{2}$MAl$_{7}$Ge$_{4}$ and have an increase in site symmetry to $\bar{4}$.  Furthermore, though the Ge atoms are surrounded by both Al and Ce, the local environment around Al is remarkably complex and consists of close contacts with Ce, M, and Ge.  The complexity of the local environment surrounding the Al centers also significantly contributes to the reduction in both the site and translational symmetry.  The Al and Ge site symmetry is never greater than twofold in Ce$_{2}$MAl$_{7}$Ge$_{4}$; whereas, two of the Ge sites in  Sm$_2$NiGa$_{12}$  reside on the 2/m positions.

A Rietveld refinement of the crystal structure of Ce$_{2}$CoAl$_{7}$Ge$_{4}$ based on the neutron powder pattern collected above the magnetic transition temperature at 4.2 K is depicted in Fig. \ref{Figure1}e. The Al-peaks from the sample holder can were excluded from the refinement pattern. A similar refinement of the neutron powder pattern of Ce$_{2}$NiAl$_{7}$Ge$_{4}$ collected at 2 K was also performed. A crystal structure with space group P\={4}2$_{1}$m determined by single crystal x-ray diffraction as described above was used in the refinement in each case. The thermal parameters were held fixed to 0 assuming that the Debye Waller factor at the measured temperatures is negligible. The results of the refinement are presented in Table \ref{Table-neutron}. In each sample, a few small peaks were not successfully indexed (as marked by asterisks in the Fig. \ref{Figure1}e) and are attributed to a small amount of an unknown impurity phase. 

\begin{table}[h]
\caption{Crystallographic parameters and agreement factors from a Rietveld refinement of the neutron powder pattern of Ce$_{2}$CoAl$_{7}$Ge$_{4}$ and Ce$_{2}$NiAl$_{7}$Ge$_{4}$ in space group P\={4}2$_{1}$m with atomic coordinates given in Tables \ref{Table-Co} and \ref{Table-Ni}, respectively.}\label{Table-neutron}
\begin{center}
\begin{tabular}{l@{\hspace{0.7cm}}c@{\hspace{0.7cm}}c}
\hline
	                    &   Ce$_{2}$CoAl$_{7}$Ge$_{4}$	&	Ce$_{2}$NiAl$_{7}$Ge$_{4}$	\\
\hline					
T (K)       	        &	4.2  	                &	2.0 	                \\
$\lambda$ (\AA)           & 2.0775                    &   2.0775              \\
a (\AA)	                &	5.9053(8)           &	5.9188(2)           \\
c (\AA)	                &	15.3314(8)	            &	15.2354(7)	             \\
Bragg R-factor	        &	4.5                &	6.3                   \\
$\chi^{2}$	            &	4.0	                &	4.8                    \\
\hline					
\end{tabular}
\end{center}
\end{table}

\subsection{Magnetic Properties}
\begin{figure}[h]
\begin{center}
\includegraphics[scale=1]{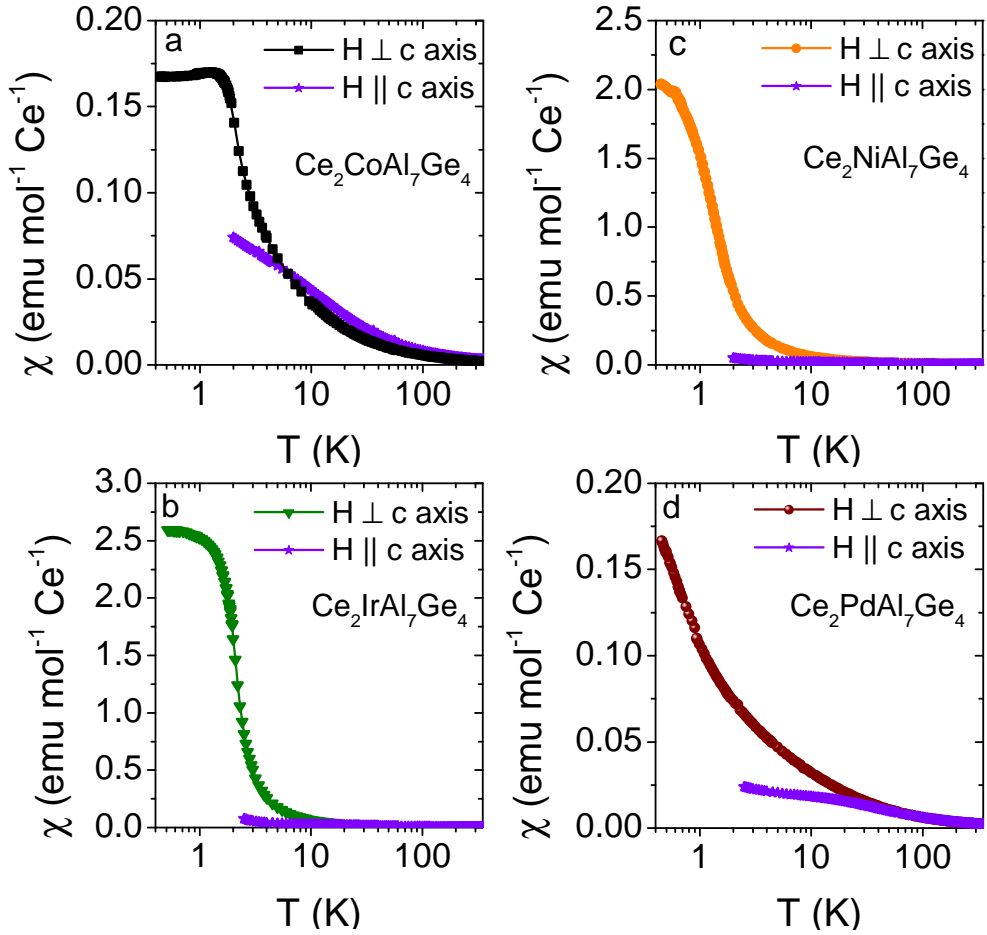}
\caption{(Color online) Magnetic susceptibility $\chi$ of Ce$_{2}$MAl$_{7}$Ge$_{4}$ (M=Co, Ni, Ir, Pd) measured in an external magnetic field of $H$ = 1.1 kOe applied parallel and perpendicular to the crystallographic \emph{c}-axis.}\label{F2}
\end{center}
\end{figure}
\begin{figure}[h]
\begin{center}
\includegraphics[scale=.9]{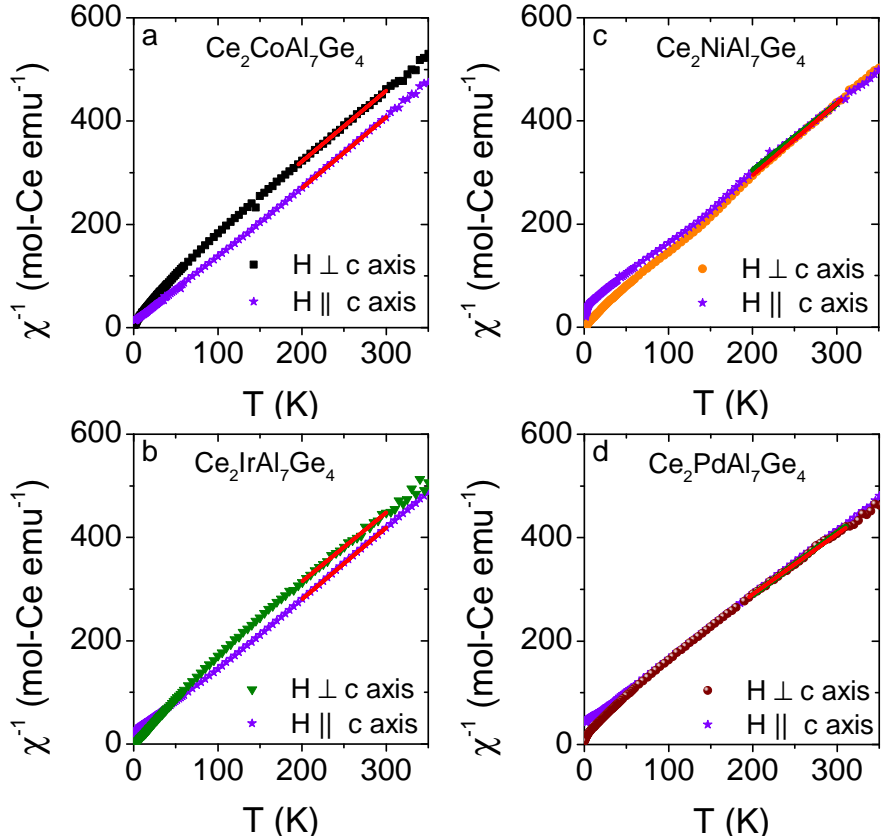}
\caption{(Color online) Inverse magnetic susceptibility $\chi^{-1}$(T) of Ce$_{2}$MAl$_{7}$Ge$_{4}$ (M=Co, Ni, Ir, Pd) measured in an external magnetic field of $H$ = 1.1 kOe applied parallel and perpendicular to the crystallographic \emph{c}-axis. The solid line is a Curie-Weiss fit to the data.}\label{F2b}
\end{center}
\end{figure}

The magnetic susceptibility  $\chi \equiv  M/H$ for $H$ = 1.1 kOe applied parallel and perpendicular to the crystallographic \emph{c}-axis of  Ce$_{2}$MAl$_{7}$Ge$_{4}$ (M=Co, Ni, Ir, Pd) is displayed in Figs. \ref{F2}a-d. For all of the compounds, below 5 K, $\chi$ measured perpendicular to the \emph{c}-axis is greater than that measured parallel to the \emph{c}-axis, indicating easy plane magnetic moments.  The high temperature susceptibility shows Curie-Weiss behavior in all four compounds. A Curie-Weiss fit to the data of the form $\chi$ =  $C/(T-\theta_{CW})$, where $C$ = $N_{A}\mu^{2}_{eff}/3k_{B}$ is the Curie constant, $N_{A}$ is Avogadro's number and $k_{B}$ is the Boltzmann constant, shown in Figs. \ref{F2b}a-d, yields the effective moment $\mu_{eff}$ and Curie-Weiss temperature $\theta_{CW}$, which are presented in Table \ref{Table-Curie}. The polycrystalline average effective moment for  Ce$_{2}$CoAl$_{7}$Ge$_{4}$, Ce$_{2}$IrAl$_{7}$Ge$_{4}$, Ce$_{2}$NiAl$_{7}$Ge$_{4}$ and Ce$_{2}$PdAl$_{7}$Ge$_{4}$ is 2.42, 2.42, 2.43, and 2.55 $\mu_{B}$/Ce, consistent with Ce being in a Ce$^{3+}$ state for which the expected moment is 2.54 $\mu_{B}$. Moreover, these effective moments indicate that the transition metal M ions  are nonmagnetic in Ce$_{2}$MAl$_{7}$Ge$_{4}$ ($M$=Co, Ni, Ir, Pd).  In all of the compounds, the Curie-Weiss temperatures are negative, indicating overall antiferromagnetic (and/or Kondo) interactions. Ce$_{2}$CoAl$_{7}$Ge$_{4}$, Ce$_{2}$NiAl$_{7}$Ge$_{4}$, and Ce$_{2}$IrAl$_{7}$Ge$_{4}$ show a tendency towards saturation below 2 K, indicating that these compounds order magnetically.  Given the large value of $\chi$ ($>$ 2 emu/mol-Ce) at 0.45 K in the ordered state for the M=Ni and Ir compounds, these materials likely order in a ferromagnetic-like state.  For Ce$_{2}$CoAl$_{7}$Ge$_{4}$ and Ce$_{2}$PdAl$_{7}$Ge$_{4}$, $\chi$T vs T decreases with decreasing temperature down to 0.45 K (not shown), suggesting the presence of antiferromagnetic correlations.
\begin{table}[h]
\caption{Effective magnetic moment $\mu_{eff}$ and Curie-Weiss temperature $\theta_{CW}$ of Ce$_{2}$MAl$_{7}$Ge$_{4}$  compounds obtained from a Curie-Weiss fit to the high temperature inverse susceptibility $\chi^{-1}$.}\label{Table-Curie}
  \begin{center}
  \begin{tabular}{lcccc}
\hline								
Compound	                                       &	Ce$_{2}$CoAl$_{7}$Ge$_{4}$ 	&	Ce$_{2}$IrAl$_{7}$Ge$_{4}$ 	& Ce$_{2}$NiAl$_{7}$Ge$_{4}$  & Ce$_{2}$PdAl$_{7}$Ge$_{4}$  \\
\hline							
$\mu_{eff}$ (H $ \parallel$ \emph{c}-axis) ($\mu_{B}$)   & 2.41$^{\dag}$                   & 2.41         &  2.50         &  2.54   \\
$\mu_{eff}$ (H $\perp$ \emph{c}-axis)	($\mu_{B}$)      & 2.42                      & 2.42       &  2.39         &  2.56   \\
$\mu_{eff}$ (poly. avg.) ($\mu_{B}$)                     & 2.42         	        &  2.42       &  2.43         &  2.55  \\
$\theta_{CW}$ H $ \parallel$ \emph{c}-axis (K)           & 3$^{\dag}$ 	                & -3  	  & -38        & -34       \\
$\theta_{CW}$ H $\perp$ \emph{c}-axis (K)	             & -37  	                & -30  	  & -11        & -43 	      \\
$\theta_{CW}$(poly. avg.) (K)                            &-17                    &  -21     & -20        & -40         \\
\hline
\end{tabular}
\\
$\dag$ \footnotesize{A modified Curie-Weiss law of the form $\chi$ =  $\chi_{0}$ + $C/(T-\theta_{CW})$ was used to determine these values.}
\end{center}
\end{table}
\begin{figure}[H]
\begin{center}
\includegraphics[scale=1]{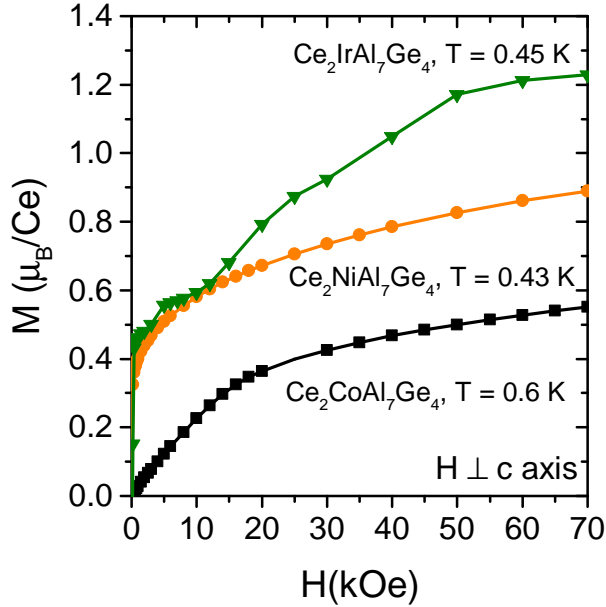}
\caption{(Color online) $M$ vs. $H$ of Ce$_{2}$CoAl$_{7}$Ge$_{4}$, Ce$_{2}$IrAl$_{7}$Ge$_{4}$ and Ce$_{2}$NiAl$_{7}$Ge$_{4}$, measured in the magnetically ordered state at the temperatures specified.}\label{F3}
\end{center}
\end{figure}

The $M$ vs. $H$ in the magnetically ordered state of Ce$_{2}$CoAl$_{7}$Ge$_{4}$, Ce$_{2}$NiAl$_{7}$Ge$_{4}$, and Ce$_{2}$IrAl$_{7}$Ge$_{4}$ is shown in Fig. \ref{F3}. $M(H)$ for Ce$_{2}$CoAl$_{7}$Ge$_{4}$  is  linear up to 20 kOe and then bends over at higher fields, although it does not completely saturate up to 70 kOe. In Ce$_{2}$NiAl$_{7}$Ge$_{4}$ and Ce$_{2}$IrAl$_{7}$Ge$_{4}$, $M(H)$ exhibits a small hysteresis ($\sim$150 Oe for M=Ni and $\sim$400 Oe for M=Ir), characteristic of a soft ferromagnet. A ferromagnetic-like state is consistent with a large value of $\chi$(T) below 5 K for M=Ni, Ir, and is nearly an order of magnitude larger than that of Ce$_{2}$CoAl$_{7}$Ge$_{4}$ and Ce$_{2}$PdAl$_{7}$Ge$_{4}$. Furthermore, a plot of $\chi$T vs T (not shown) increases rapidly with decreasing temperature just above the ordering temperature, indicating the presence of ferromagnetic correlations in Ce$_{2}$MAl$_{7}$Ge$_{4}$ (M=Ni, Ir).   At higher fields, the hysteresis is negligible in Ce$_{2}$IrAl$_{7}$Ge$_{4}$, and there are plateaus at 8, 16 and 24 kOe and a slope change at 50 kOe.   The magnetic properties of Ce$_{2}$MAl$_{7}$Ge$_{4}$ (M=Co, Ni, Ir) in magnetic field do not show typical behavior expected for a simple ferromagnet or antiferromagnet, suggesting these materials exhibit some form of complex magnetic ordering below 2 K, perhaps in an incommensurate magnetic state.  This is consistent neutron powder diffraction measurements well below $T_M$ of Ce$_{2}$CoAl$_{7}$Ge$_{4}$ and Ce$_{2}$NiAl$_{7}$Ge$_{4}$, which found no evidence for additional peaks or extra intensity corresponding to magnetic ordering, within the instrument resolution.  The magnetic moment per Ce in  Ce$_{2}$CoAl$_{7}$Ge$_{4}$, Ce$_{2}$IrAl$_{7}$Ge$_{4}$ and Ce$_{2}$NiAl$_{7}$Ge$_{4}$ at the highest field measured of 70 kOe (Fig. \ref{F3}) is 0.55, 1.2, and 0.9 $\mu_{B}$, respectively. These values are much smaller than the  free Ce$^{3+}$ ion moment, given by $g_{J}J$ = 2.14 $\mu_{B}$.  A reduction of the ordered moment in  rare earth systems may be due to crystal field effects and/or Kondo screening. Ce$_{2}$PdAl$_{7}$Ge$_{4}$ does not show any sign of magnetic ordering down to the lowest temperature measured (0.46 K).

\subsection{Specific Heat}

\begin{figure}[H]
\begin{center}
\includegraphics[scale=.55]{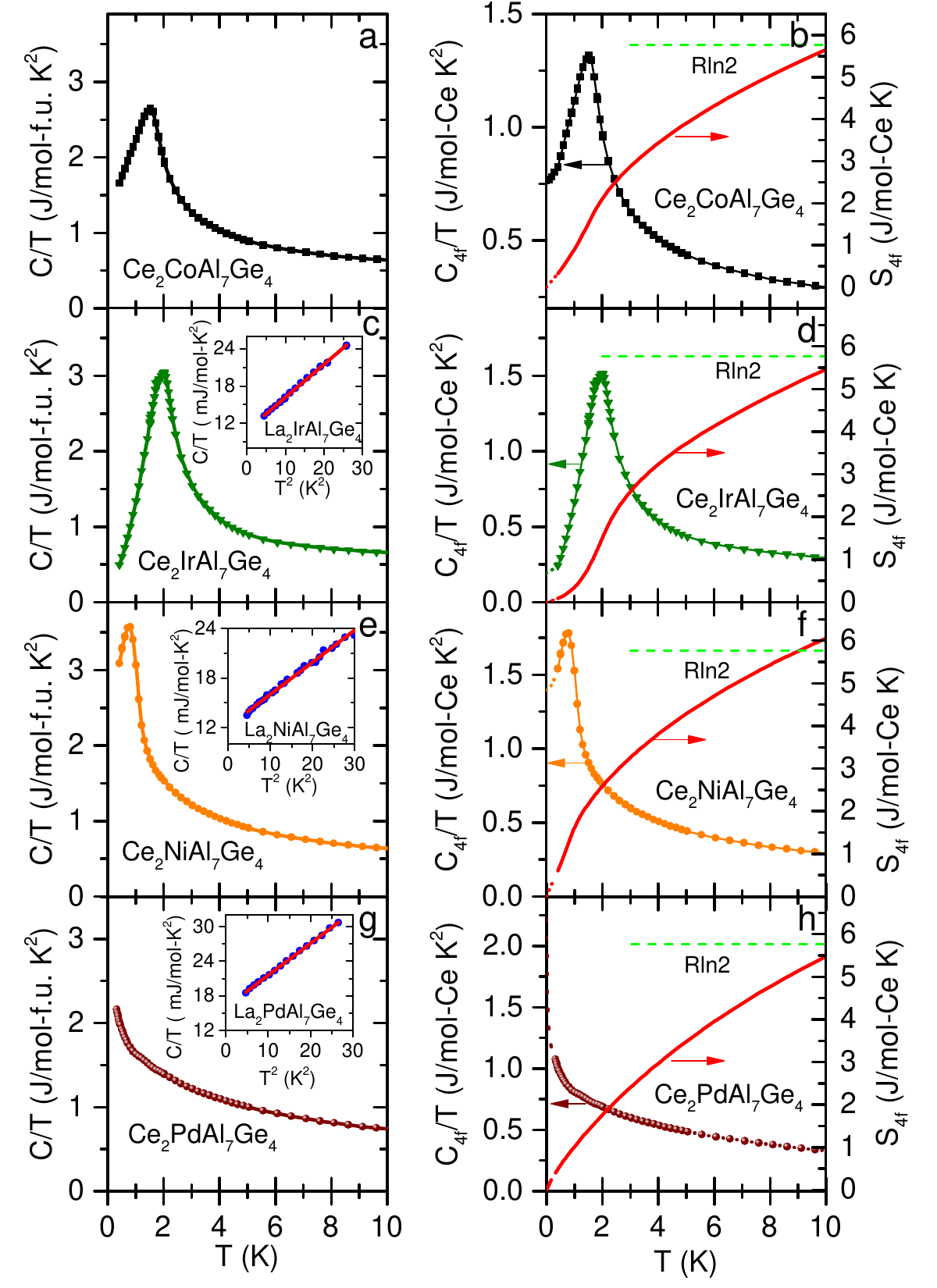}
\caption{(Color online) Specific heat divided by temperature $C/T$ as a function of temperature of a) Ce$_{2}$CoAl$_{7}$Ge$_{4}$, c) Ce$_{2}$IrAl$_{7}$Ge$_{4}$, e) Ce$_{2}$NiAl$_{7}$Ge$_{4}$, g) Ce$_{2}$PdAl$_{7}$Ge$_{4}$. On the scale of these plots, $C/T$ of the corresponding non-magnetic, isostructural  La$_{2}$MAl$_{7}$Ge$_{4}$ analogs is negligibly small. The insets show the low temperature fit of the heat capacity of the La-analogs. The  right panels show the  $4f$ contribution to the specific heat $C_{4f}/T$ (left axis) and temperature evolution of the $4f$ contribution to the entropy $S_{4f}$ (right axis) of b) Ce$_{2}$CoAl$_{7}$Ge$_{4}$, d) Ce$_{2}$IrAl$_{7}$Ge$_{4}$, f) Ce$_{2}$NiAl$_{7}$Ge$_{4}$, h) Ce$_{2}$PdAl$_{7}$Ge$_{4}$. The dotted lines are an extrapolation of \emph{C$_{4f}$/T} to $T=0$ and are used to estimate the entropy below 0.4 K.}\label{F4}
\end{center}
\end{figure}

The specific heat, plotted as $C/T$, for the Ce$_{2}$MAl$_{7}$Ge$_{4}$ compounds is shown in Figs. \ref{F4}(a,c,e,g). The $4f$ contribution to the heat capacity, plotted in Figs. \ref{F4}(b,d,f,h) as $C_{4f}/T$ is obtained by subtracting $C/T$ of La-counterparts La$_{2}$MAl$_{7}$Ge$_{4}$ from that of Ce$_{2}$MAl$_{7}$Ge$_{4}$. The $4f$ contribution to the entropy ($S_{4f}$) obtained by integrating $C_{4f}/T$ with respect to $T$, also is shown in Figs. \ref{F4}(b,d,f,h). The contribution of La$_{2}$NiAl$_{7}$Ge$_{4}$ was used to determine $C_{4f}/T$ and $S_{4f}$ of Ce$_{2}$CoAl$_{7}$Ge$_{4}$ because crystals of La$_{2}$CoAl$_{7}$Ge$_{4}$ were not grown.  The La-compounds show a behavior typical of a simple metal. The electronic contribution to the specific heat $\gamma$, obtained from a linear fit to \emph{C/T} vs. $T^{2}$ below 5.5 K of the form \emph{C/T} = $\gamma$ + $\beta T^2$ [insets in Fig. \ref{F4}(c,e,g)], are 10.8, 12.1 and 16.1 (mJ/mol-f.u.-K$^{2}$) for La$_{2}$IrAl$_{7}$Ge$_{4}$, La$_{2}$NiAl$_{7}$Ge$_{4}$ and  La$_{2}$PdAl$_{7}$Ge$_{4}$, respectively. 
The respective phonon specific heat coefficient $\beta$ is 0.53, 0.51 and 0.40 (mJ/mol-f.u.-K$^{4}$). The Debye temperature estimated from $\beta$ using the relation $\theta_{D}$ =$\sqrt[3]{12\pi^{4}rR/5\beta}$ (\emph{r} is number of atoms in the formula unit) for La$_{2}$IrAl$_{7}$Ge$_{4}$, La$_{2}$NiAl$_{7}$Ge$_{4}$ and  La$_{2}$PdAl$_{7}$Ge$_{4}$ is 372, 412 and 367 K, respectively. A peak is observed in $C/T$ signaling the onset of magnetic order at $T_M$= 1.5, 2.0, and 0.8 K in Ce$_{2}$CoAl$_{7}$Ge$_{4}$, Ce$_{2}$IrAl$_{7}$Ge$_{4}$, and Ce$_{2}$NiAl$_{7}$Ge$_{4}$, respectively.   No   peak is observed in Ce$_{2}$PdAl$_{7}$Ge$_{4}$ down to 0.4 K. Instead $C/T$ increases with decreasing temperature down to 0.4 K where  it reaches a value of $\gamma$ = 1.0 J/mol-Ce-K$^{2}$.

The entropy at the magnetic transition  $S_{4f}$($T_M$) (per Ce) of Ce$_{2}$CoAl$_{7}$Ge$_{4}$, Ce$_{2}$IrAl$_{7}$Ge$_{4}$, and Ce$_{2}$NiAl$_{7}$Ge$_{4}$ is 0.27, 0.23 and 0.22 of $Rln2$, but the entropy of the CEF doublet ground state is recovered by $\sim$ 10 K in the  Ce$_{2}$MAl$_{7}$Ge$_{4}$ compounds [Fig. \ref{F4}(b,d,f,h)]. Such a strong reduction of  $S_{4f}$ at $T_M$ is expected when Kondo screening of the Ce moments occurs. A simple model relates the reduction of the specific heat jump at the magnetic transition $\Delta C (T_{M})$ to the ratio of Kondo and ordering temperatures $T_K/T_{mag}$.\cite{Besnus1992}  From a linear extrapolation of $C/T$ from 10 K to $T_M$ [Fig. \ref{F4}(b,d,f,h)], the obtained peak heights  $\Delta C \sim$ 1.0, 1.6, and 1.0 J/mol-Ce-K for Ce$_{2}$MAl$_{7}$Ge$_{4}$, M=Co, Ir, Ni, respectively, are significantly reduced from the expected  $\Delta C = 12.48$ J/mol-Ce-K for a $J=1/2$ two-level magnetic system.\cite{Besnus1992}  From these values of  $\Delta C$, we estimate that $T_K/T_{mag} \sim$ 5, 3, 5 for M=Co, Ir, Ni, respectively, yielding $T_K \sim$ 5, 8, and 5 K.  These values of $T_K$ are in accord with the fact that the $4f$ entropy reaches $(1/2)Rln2$ around 5 K in Ce$_{2}$MAl$_{7}$Ge$_{4}$, as expected from the calculations of Rajan.\cite{Rajan1983} Because  $C_{4f}/T$ increases over a broad temperature range as the magnetic transition is approached in the paramagnetic state, it is not possible to estimate accurately the electronic contribution to the specific heat ($\gamma$). Crudely, however, $\gamma \sim Rln2/T_{K}$, which implies that these are heavy fermion systems with a Kondo temperature $T_K \sim 5$ K.

\subsection{Electrical Resistivity}

\begin{figure}[h]
\begin{center}
\includegraphics[scale=.8]{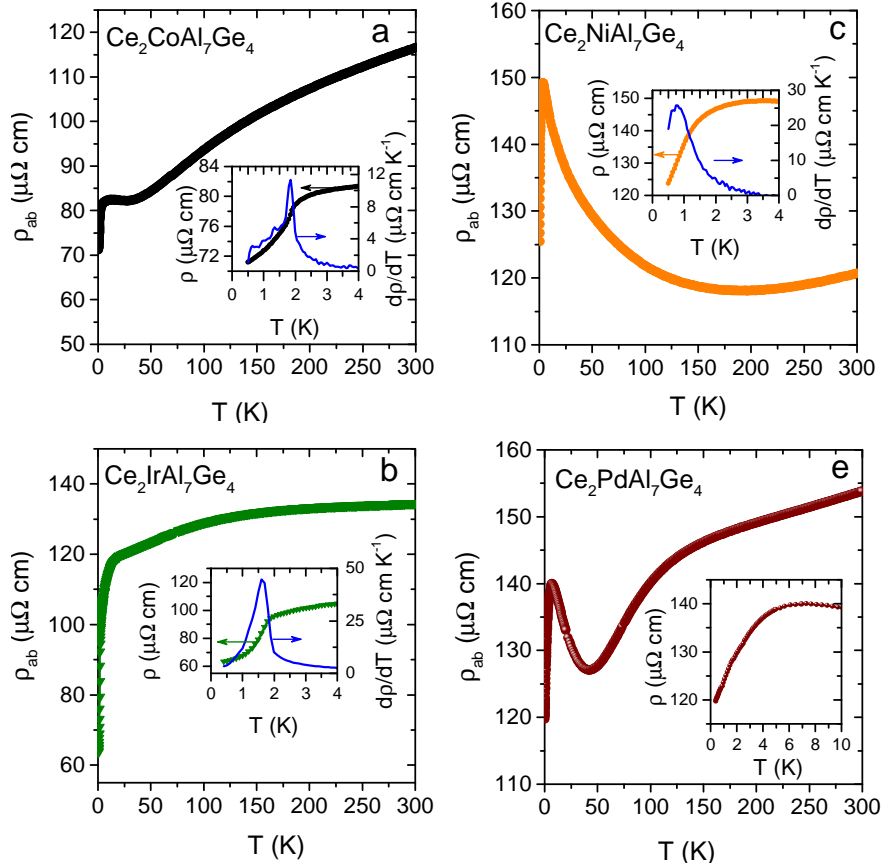}
\caption{(Color online) Electrical resistivity  as a function of temperature $\rho(T)$ with current applied in the $ab$-plane of a) Ce$_{2}$CoAl$_{7}$Ge$_{4}$ , b) Ce$_{2}$IrAl$_{7}$Ge$_{4}$, c) Ce$_{2}$NiAl$_{7}$Ge$_{4}$, and d) Ce$_{2}$PdAl$_{7}$Ge$_{4}$. The insets in (a - c) show the data between 0.4 and 4 K near the magnetic transition, along with the temperature derivative of the resistivity $d\rho/dT$. The inset in (d) shows the electrical resistivity of Ce$_{2}$PdAl$_{7}$Ge$_{4}$ between 0.4  and 10 K. The resistivity of Ce$_{2}$CoAl$_{7}$Ge$_{4}$ and Ce$_{2}$NiAl$_{7}$Ge$_{4}$ was measured in a small external magnetic field of 200 Oe and 500 Oe, respectively.}\label{F5}
\end{center}
\end{figure}

\begin{figure}[h]
\begin{center}
\includegraphics[scale=.8]{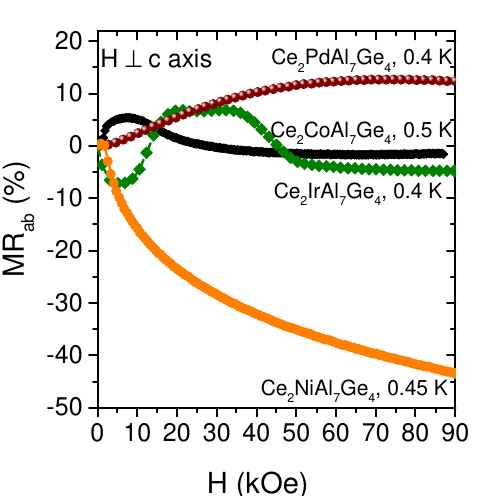}
\caption{(Color online) Transverse magnetoresistance MR of Ce$_{2}$MAl$_{7}$Ge$_{4}$ with the current and the external magnetic field applied within the $ab$-plane measured at the temperature specified.}\label{F6}
\end{center}
\end{figure}

Figure \ref{F5} shows the electrical resistivity of the Ce$_{2}$MAl$_{7}$Ge$_{4}$ compounds measured with current applied in the \emph{ab}-plane. In all four compounds, the electrical resistivity is weakly temperature dependent above 150 K. With decreasing temperature in Ce$_{2}$CoAl$_{7}$Ge$_{4}$, Ce$_{2}$IrAl$_{7}$Ge$_{4}$ and  Ce$_{2}$PdAl$_{7}$Ge$_{4}$,   a broad ``knee'' appears in $\rho(T)$ around 100-150 K.  Below 100 K,  $\rho(T)$ of Ce$_{2}$PdAl$_{7}$Ge$_{4}$ passes through a distinct minimum at 45 K and a peak at 7 K, below which it drops sharply. Ce$_{2}$CoAl$_{7}$Ge$_{4}$ also shows a shallow minimum at 30 K and a peak at 10 K. In Ce$_{2}$IrAl$_{7}$Ge$_{4}$, the resistivity drops sharply below 20 K. Ce$_{2}$NiAl$_{7}$Ge$_{4}$ shows a shallow resistivity minimum at relatively higher temperature (200 K), below which resistivity gradually increases making a peak at 4 K with a sharp drop below it. The three Ce$_{2}$CoAl$_{7}$Ge$_{4}$, Ce$_{2}$IrAl$_{7}$Ge$_{4}$, and Ce$_{2}$NiAl$_{7}$Ge$_{4}$ materials show a kink at the ordering temperature at $T_M$ =  1.8, 1.8, and 0.8 K, respectively (inset of Figs. \ref{F5}a,b,c), in reasonable agreement with specific heat and magnetic susceptibility measurements.  This feature is also reflected in the  temperature derivative of resistivity $d\rho/dT$ as a peak (inset of Figs. \ref{F5}a,b,c). No such feature is observed in  Ce$_{2}$PdAl$_{7}$Ge$_{4}$ as expected from the susceptibility and heat capacity data.

The magnetoresistance (MR) defined as $[(\rho_{H}-\rho_{0})/\rho_{0}]\times100 \%$ is depicted in Fig. \ref{F6} as a function of external magnetic field for the Ce$_{2}$MAl$_{7}$Ge$_{4}$ compounds measured with both the magnetic field and current applied in the \emph{ab}-plane in a transverse geometry.  The MR of Ce$_{2}$CoAl$_{7}$Ge$_{4}$ at T=0.6 K first increases with increasing magnetic field and shows a small jump at 2 kOe, goes through a maximum at 8.5 kOe and then decreases with increasing $H$. It becomes negative at 28 kOe and saturates above 60 kOe. These features observed in MR correspond to features observed in $M$ vs. $H$ measurements (Fig. \ref{F3}). Although not clear in  $M$ vs. $H$, $dM$/$dH$ vs. $H$ (not shown) shows a decrease at 2 and 15 kOe. This MR behavior indicates that Ce$_{2}$CoAl$_{7}$Ge$_{4}$ does not have a metamagnetic transition up to 87 kOe. The MR of Ce$_{2}$IrAl$_{7}$Ge$_{4}$ at 0.4 K is more dramatic with features at 7, 19, 36, and 50 kOe that closely follow those observed in $M$ vs. $H$ (Fig. \ref{F3}). At high fields, the MR is negative and saturates at -5\% at 90 kOe. In Ce$_{2}$NiAl$_{7}$Ge$_{4}$, the magnetoresistance at 0.45 K decreases monotonically with increasing field up to 90 kOe where it reaches a very large negative value of 43 \%. This behavior suggests that there are significant magnetic fluctuations in Ce$_{2}$NiAl$_{7}$Ge$_{4}$  at 0.45 K that are suppressed with a magnetic field, resulting in a large negative magnetoresistance. In contrast, the MR of Ce$_{2}$PdAl$_{7}$Ge$_{4}$ measured at 0.4 K increases with increasing $H$ up to 70 kOe, then decreases slightly to 12.3\% at  90 kOe.

\subsection{Crystalline Electric Field Analysis of  Ce$_{2}$PdAl$_{7}$Ge$_{4}$}
\begin{figure}[ht]
\begin{center}
\includegraphics[scale=.8]{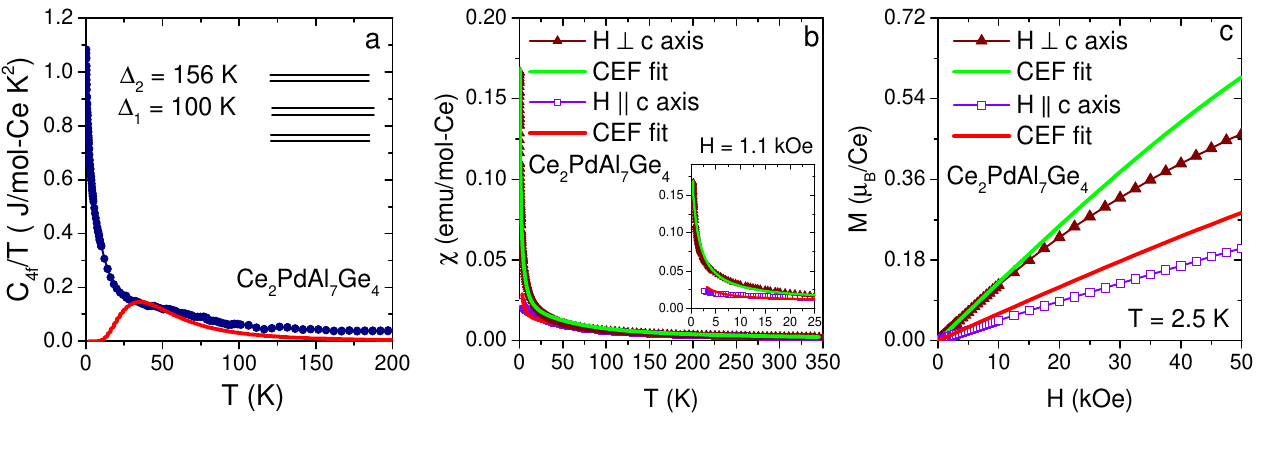}
\caption{(Color online) (a) Fits of a CEF model to the a) heat capacity, b) magnetic susceptibility and c) magnetization data of Ce$_{2}$PdAl$_{7}$Ge$_{4}$. Inset in (b) shows $\chi$ vs. T below 25 K.}\label{F4b}
\end{center}
\end{figure}

To better understand the interesting properties of  Ce$_{2}$PdAl$_{7}$Ge$_{4}$ as well as to shed light on its crystal field scheme, we have analyzed the data presented in  Fig. \ref{F4b} using model including interactions between nearest-neighbors as well as the orthorhombic CEF hamiltonian:
\begin{equation}\label{equation1}
\mathcal{H}=J_{AFM}\sum_{<i,l>}j_{i}\cdot j_{l} + B_{2}^{0}O_{2}^{0} + B_{2}^{2}O_{2}^{2}+ B_{4}^{0}O_{4}^{0} + B_{4}^{2}O_{4}^{2} + B_{4}^{4}O_{4}^{4},
\end{equation}
\noindent where $J_{AFM}>0$ represents an AFM interaction between nearest neighbor local spins $j_{i}$, $B_{i}^{n}$ are the CEF parameters, and $O_{i}^{n}$ are the Stevens equivalent operators obtained from the angular momentum operators.\cite{Hutchings1964, Stevens1952} The first term in the Hamiltonian in Eq. \ref{equation1} may be simplified to $zJj\cdot <j>$ via a mean field approximation ($j_{i}\cdot j_{l} \sim j\cdot <j>$), where $z$ is the number of nearest neighbors. Although $\mathcal{H}$ may be solved by this approach, the best fits to the data are achieved when  a second antiferromagnetic exchange interaction $J'_{AFM}$ is included, which agrees with the results from the Curie-Weiss fits at high temperatures. This model was then used to simultaneously fit $\chi(T)$ in the whole temperature range, $M(H)$ at $2.5$~K  and $C/T$ with $T> 16$ K as a constraint, yielding the CEF parameters as well as the antiferromagnetic RKKY exchange parameters displayed in Table \ref{TableCEF}. When $B_{2}^{0}>0$, the Stevens operator $O_{2,i}^{0} = 3\hat{J}_{z,i}^{2} - J(J+1)$ favors in-plane alignment of spins (i.e., $\hat{J}_{z} = 0$), which is consistent with the magnetic susceptibility data ($\chi_{ab} > \chi_{c}$). In fact,  there is good agreement between the fits and the experimental $\chi$(T) data (Fig. \ref{F4b}b). The $M(H)$ data at $2.5$~K, however, cannot be exactly reproduced, as shown in Fig. \ref{F4b}c, although the anisotropy and the shape of the magnetization fits are  reasonable. The observed discrepancy likely arises from the fact that the model is an approximation that takes into account only single ion CEF effects and mean field interactions, which mimic RKKY interactions, and does not include any additional effects from the Kondo interactions and/or quantum fluctuations. Thus, although there is good agreement at high temperatures, the CEF model should not be expected to account for the low temperature properties of Ce$_{2}$PdAl$_{7}$Ge$_{4}$ discussed below.

\begin{table*}
\caption{Extracted crystal field parameters (in Kelvin) for Ce$_{2}$PdAl$_{7}$Ge$_{4}$. Here, $z_{AFM}$ is the number of nearest neighbors with AFM coupling.}\label{TableCEF}
\begin{centering}
\begin{tabular}{|c@{\hspace{0.6cm}}||c@{\hspace{0.6cm}}|c@{\hspace{0.6cm}}|c@{\hspace{0.6cm}}|c@{\hspace{0.6cm}}|c@{\hspace{0.6cm}}|c@{\hspace{0.6cm}}|c@{\hspace{0.6cm}}|}
\hline
 &  B$^{0}_{2}$ &  B$^{2}_{2}$ & B$^{0}_{4}$ &  B$^{4}_{2}$ & B$^{4}_{4}$ & $z_{AFM}J_{AFM}$ & $z'_{AFM}J'_{AFM}$ \tabularnewline
\hline
Ce$_{2}$PdAl$_{7}$Ge$_{4}$ &   1.79   &  -0.53 & 0.24  &  0.016  & 2.49  & 0.87   &  0.52        \tabularnewline
\hline
\end{tabular}
\par\end{centering}
\end{table*}

The extracted parameters resulted in the CEF level scheme shown in Fig. \ref{F4b}a for Ce$_{2}$PdAl$_{7}$Ge$_{4}$ and is summarized in Table \ref{TableCEF2}. The ground state doublet is separated from the first excited state and the second excited state by 100 K and 156 K, respectively. We note that the obtained CEF scheme can reproduce the main features of the data shown in Fig. \ref{F4b}, i.e., the Schottky anomaly and the magnetic anisotropy. A broad feature in the electrical resistivity  at $\sim$ 125 K is also consistent with scattering associated with depopulation of the excited CEF levels.  If this is the case, then a similar CEF scheme might be expected for Ce$_{2}$CoAl$_{7}$Ge$_{4}$ and Ce$_{2}$IrAl$_{7}$Ge$_{4}$ (although it is less clear in Ce$_{2}$NiAl$_{7}$Ge$_{4}$). In addition, the exchange interactions are small ($<$ 1 K) and suggest the presence of antiferromagnetic interactions at low temperature.  Nevertheless, it is important to emphasize that the CEF parameters obtained from fits to macroscopic measurements may not be unique and/or extremely precise. An accurate determination of the CEF scheme and its parameters requires a direct measurement by, for example, inelastic neutron scattering,\cite{Goremychkin1993} or the mixing parameters of the ground state wavefunction may be obtained from X-ray absorption studies.\cite{Willers2010}

 \begin{table*}[!ht]
  \caption{Energy levels and wave functions of the CEF scheme obtained from the thermodynamic properties of Ce$_{2}$PdAl$_{7}$Ge$_{4}$.}\label{TableCEF2}
  \centering
  \begin{tabular}{c@{\hspace{0.7cm}}c@{\hspace{0.7cm}}c@{\hspace{0.7cm}}c@{\hspace{0.7cm}}c@{\hspace{0.7cm}}c@{\hspace{0.7cm}}c@{\hspace{0.7cm}}}
\hline
\multicolumn{7}{c}{Energy levels and wave functions} \\
\cline{1-7}\\
$E (K)$  & $|-5/2\rangle$ & $|-3/2\rangle$ & $|-1/2\rangle$ & $|+1/2\rangle$ & $|+3/2\rangle$ & $|+5/2\rangle$ \\
&&&&&&\\
156  & -0.87   &   -0.015  &   0.041 &  0.0012 &    -0.495 &   -0.025 \\
156 & -0.025  &  0.495 &  0.0012 &  -0.041 &     0.015  &   0.87 \\
100 & -0.006 & 0.006 & -0.13 & 0.99 & 0.0 & 0.04 \\
100  & 0.04  &  0.0  &  0.99 &   0.13  & 0.006 &  0.006 \\
0     & -0.089  &   0.85  &  0.003 &  0.016 &    0.16  &   -0.49 \\
0     & -0.49   &  -0.16  &   0.016  &  -0.003  &     0.85 &   0.089 \\
\hline
\end{tabular}
 \end{table*}

\section{Discussion}

 \begin{figure}[ht]
\begin{center}
\includegraphics[scale=1]{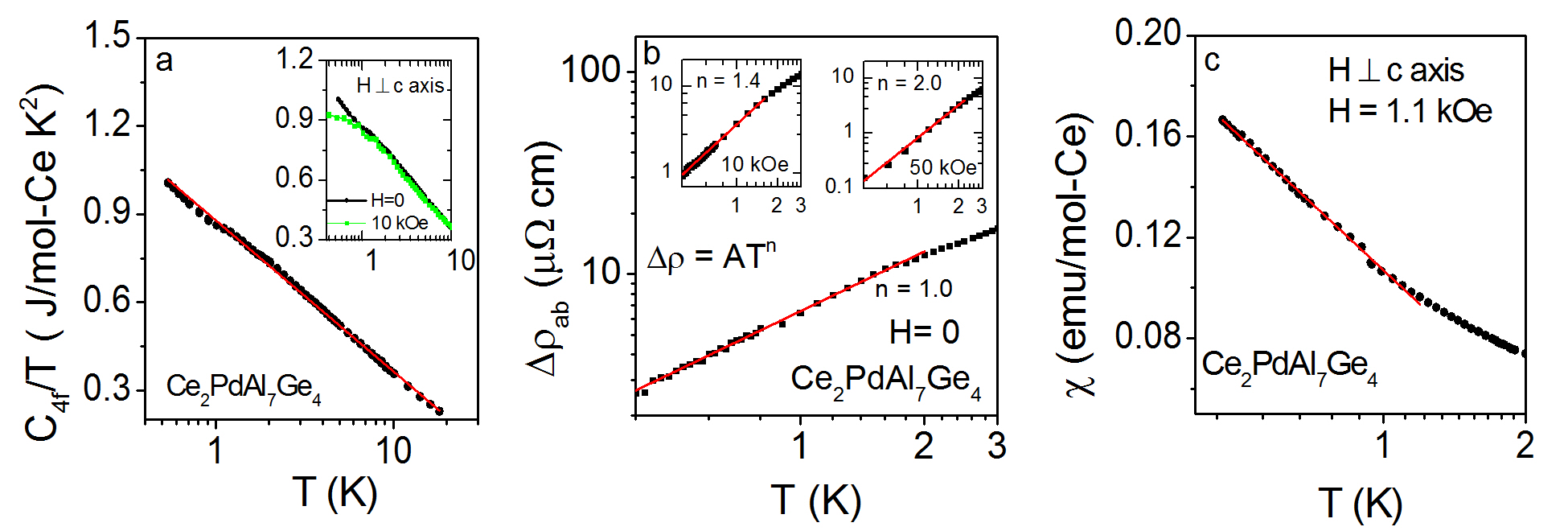}
\caption{(Color online) (a) $C_{4f}/T$  vs $T$ of Ce$_{2}$PdAl$_{7}$Ge$_{4}$ on a semi-log plot. The inset shows $C_{4f}/T$ measured in H =0 and 10 kOe, applied parallel to the $ab$-plane.  (b) log-log plot of $\triangle\rho$  vs T of Ce$_{2}$PdAl$_{7}$Ge$_{4}$ between 0.4 and 3 K for H = 0 (main panel), 10 kOe (left inset) and 50 kOe (right inset).  $\triangle\rho$ = $\rho$ - $\rho_{0}$, $\rho_{0}$ is the extrapolated $\rho$  at T = 0. The solid lines are the power law ($\triangle\rho$ = AT$^{n}$) fits to the data. The magnetic field was applied parallel to the $ab$-plane. (c) $\chi$ vs $T$ of Ce$_{2}$PdAl$_{7}$Ge$_{4}$ on a semi-log plot. In the main panels,  the solid line is a linear fit to the  data. }\label{F7}
\end{center}
\end{figure}

Among the four compounds studied Ce$_{2}$CoAl$_{7}$Ge$_{4}$,  Ce$_{2}$NiAl$_{7}$Ge$_{4}$, Ce$_{2}$IrAl$_{7}$Ge$_{4}$, and Ce$_{2}$PdAl$_{7}$Ge$_{4}$, the first three show clear magnetic ordering below 2 K in susceptibility, resistivity and heat capacity measurements. The magnetic structure of these compounds appears to be complex, as suggested by the lack of any additional signal of magnetic order in powder neutron diffraction below $T_M$ in the M=Co, and Ni compounds and the multiple features in the $M(H)$ curve in the ordered state of the M=Ir compound.  The apparent ferromagnetic-like and antiferromagnetic ordered states in Ce$_{2}$MAl$_{7}$Ge$_{4}$ (M=Co, Ni, Ir) suggest a finely tuned balance of competing magnetic exchange interactions. It is possible that these compounds have an incommensurate magnetic structure, or that the ordered moment is too small to be detected within the resolution of the neutron powder diffraction experiment.  The low ordering temperatures, the reduced specific heat jumps at $T_M$, and the small amount of entropy released below $T_M$ ($\sim$ 25\% of $Rln2$) uniformly point to the Ce$_{2}$MAl$_{7}$Ge$_{4}$ compounds being close to a magnetic/nonmagnetic boundary in the Doniach diagram. Ce$_{2}$PdAl$_{7}$Ge$_{4}$ may be very close to this boundary as it does not show any sign of magnetic ordering down to 0.4 K. Instead, it exhibits non-Fermi liquid  behavior as shown in Fig. \ref{F7}.   $C_{4f}(T)/T$ follows a $\sim  -ln(T)$ temperature dependence between 0.4 K and 18 K (Fig. \ref{F7}a). Upon application of an in-plane magnetic field of 10 kOe, the logarithmic behavior deviates below 0.8 K (inset of Fig. \ref{F7}a).  In addition, the resistivity is linear in temperature $\rho \sim T$ between 0.4 K and 1.8 K (Fig. \ref{F7}b). In magnetic field, the low temperature resistivity deviates from $T$-linear behavior. At 10 kOe, fits of $\rho(T)$ = $\rho_{0}$ + $AT^{n}$ to the data between $0.4 < T < 1.8 $ yield $n=1.4$ (left inset in Fig. \ref{F7}b) and exhibits Fermi liquid $T^{2}$ behavior at 50 kOe below 2.4 K (right inset in Fig. \ref{F7}b).   Fig. \ref{F7}c shows the  magnetic susceptibility of Ce$_{2}$PdAl$_{7}$Ge$_{4}$ on a semi-long plot.  Between 0.4 K and 1 K, $\chi(T) \sim -ln(T)$. All these properties suggest that in the normal state Ce$_{2}$PdAl$_{7}$Ge$_{4}$ shows NFL behavior at the lowest temperature measured (0.4 K), with the application of a magnetic field  driving the system towards a Fermi liquid state.  It will be interesting to see if  Ce$_{2}$PdAl$_{7}$Ge$_{4}$ is right at the quantum critical point with non-Fermi liquid properties extending to temperatures below 0.4 K, or whether it is just away from the QCP and either orders magnetically or is a Fermi liquid at very low temperatures.  These lower temperature measurements may also allow a better understanding of the nature of the quantum fluctuations and which theoretical models\cite{vonLohneysen07,Si2001,Pepin2007,Abrahams2012}  might be appropriate for describing this system. Pressure measurements on the other magnetically ordered  Ce$_{2}$MAl$_{7}$Ge$_{4}$ compounds, in which pressure is expected to tune the system towards the magnetic/nonmagnetic boundary, would also be worthwhile.

\section{Summary}
Single crystals of the compounds Ce$_{2}$MAl$_{7}$Ge$_{4}$ (M = Co, Ir, Ni, Pd) were synthesized and their physical properties were determined by means of x-ray and neutron diffraction, magnetic susceptibility, specific heat, and electrical resistivity measurements.  These compounds crystallize in a noncentrosymmetric tetragonal space group P\={4}2$_{1}$m. Ce$_{2}$CoAl$_{7}$Ge$_{4}$, Ce$_{2}$NiAl$_{7}$Ge$_{4}$, and Ce$_{2}$IrAl$_{7}$Ge$_{4}$ order magnetically below 2 K.  Ce$_{2}$PdAl$_{7}$Ge$_{4}$ does not show magnetic ordering down to 0.4 K. An analysis of heat capacity data suggests that these materials are heavy fermion compounds with a Kondo scale of order $T_K \sim 5-8$ K that is comparable to the ordering temperatures, placing them close to a magnetic/nonmagnetic boundary. Ce$_{2}$PdAl$_{7}$Ge$_{4}$ shows NFL behavior and appears to be at, or close to, a quantum critical point.

\section{Acknowledgements}
We thank Z. Fisk, J. Lawrence and A. Mar for stimulating discussion. Work at Los Alamos National Laboratory was performed under the auspices of the US Department of Energy, Office of Basic Energy Sciences, Division of Materials Sciences and Engineering. The EDS measurements were performed at the Center for Integrated Nanotechnologies, an Office of Science User Facility operated for the U.S. Department of Energy (DOE) Office of Science. Los Alamos National Laboratory, an affirmative action equal opportunity employer, is operated by Los Alamos National Security, LLC, for the National Nuclear Security Administration of the U.S. Department of Energy under contract DE-AC52-06NA25396. P. F. S. R. acknowledges a Director's Postdoctoral Fellowship supported through the Los Alamos LDRD program. Support for T. E. A.-S. and S. K. Cary was provided by the Chemical Sciences, Geosciences, and Biosciences Division, Office of Basic Energy Sciences, Office of Science, Heavy Elements Chemistry Program, U.S. Department of Energy, under Grant DE-FG02-13ER16414. YL was supported by the Institute of Basic Sciences (IBS), Grant No. IBS-R014-D1. We acknowledge the support of the National Institute of Standards and Technology, U. S. Department of Commerce, in providing the neutron research facilities used in this work.

\bibliographystyle{ieeetr}

\begin{thebibliography}{10}
\bibitem{NIST}
Identification of any commercial product or trade name does not imply endorsement or recommendation by NIST. Unit conversions: Oe to ampere per meter (A/m)  multiply by 7.957747E+01, Molar magnetic susceptibility $½á$ = $M V_{m}/H$, where $V_{m}$ is the volume per gram mole. 1 emu/(mol Oe) = 4$\pi$ E-6 m$^{3} $/mol.
\end{thebibliography}

\pagebreak

\appendix*
\section{Atomic coordinates and equivalent displacement parameters of Ce$_{2}$MAl$_{7}$Ge$_{4}$}

\begin{table}[h]
\caption{Atomic coordinates and equivalent displacement parameters (\AA$^{2}$) for Ce$_{2}$IrAl$_{7}$Ge$_{4}$ determined by single crystal x-ray diffraction at room temperature. U$_{eq}$ is defined as one third of the trace of the orthogonalized U$^{ij}$ tensor.}\label{Table-Ir}
\begin{center}
\begin{tabular}{l@{\hspace{0.7cm}}c@{\hspace{0.7cm}}c@{\hspace{0.7cm}}c@{\hspace{0.7cm}}c@{\hspace{0.7cm}}c@{\hspace{0.7cm}}c}								
		\hline													
        Atom    & Wyck. & Occ.   &	  x	          &	      y	         &	  z	        &   U$_{eq}$ \\
		\hline						
														
         Ir(1)	&	2a	&	1	&	0	          &	0	            &	0	        &	0.005(1)	\\
         Ce(1)	&	4d	&	1	&	0	          &	0	            &	-0.2584(1)	&	0.007(1)	\\
         Ge(1)	&	2c	&	1	&	0.5	          &	0	            &	-0.2026(1)	&	0.008(1)	\\
         Ge(2)	&	2c	&	1	&	0 	          &	0.5	            &	-0.1917(1)	&	0.007(1)	\\
         Ge(3)	&	4e	&	1	&	0.2806(1)	  &	-0.2194(1)	    &	-0.4132(1)	&	0.011(1)	\\
         Al(1)	&	4e	&	1	&	-0.2946(3)	  &	-0.2054(3)	    &	-0.4277(2)	&	0.007(1)	\\
         Al(2)	&	2c	&	1	&	0	          &	-0.5	        &	-0.3555(2)	&	0.010(1)    \\
         Al(3)	&	4e	&	1	&	-0.2559(2)	  &	-0.2441(2)	    &	-0.0867(3)	&	0.008(1)	\\
         Al(4)	&	4e	&	1	&	-0.7439(2)	  &	-0.2439(2)	    &	-0.0872(3)	&	0.007(1)	\\
        \hline
        \hline
\end{tabular}
  \end{center}
\end{table}


\begin{table}[h]
\caption{Atomic coordinates and equivalent displacement parameters (\AA$^{2}$) for Ce$_{2}$NiAl$_{7}$Ge$_{4}$ determined by single crystal x-ray diffraction at room temperature. U$_{eq}$ is defined as one third of the trace of the orthogonalized U$^{ij}$ tensor.}\label{Table-Ni}
\begin{center}
\begin{tabular}{l@{\hspace{0.7cm}}c@{\hspace{0.7cm}}c@{\hspace{0.7cm}}c@{\hspace{0.7cm}}c@{\hspace{0.7cm}}c@{\hspace{0.7cm}}c}								
		\hline													
        Atom    & Wyck. & Occ.   &	  x	          &	      y	         &	  z	        &  U$_{eq}$    \\
		\hline		
        Ni(1)	&	2a	&	1	&	0	          &	0	             &	0	        &	0.005(1)	\\
        Ce(1)	&	4d	&	1	&	0	          &	0	             &	-0.2588(1)	&	0.006(1)	\\
        Ge(1)	&	2c	&	1	&	0.5	          &	0	             &	-0.1996(1)	&	0.007(1)	\\
        Ge(2)	&	2c	&	1	&	0	          &	0.5	             &	-0.1884(1)	&	0.006(1)	\\
        Ge(3)	&	4e	&	1	&	0.2798(1)	  &	-0.2202(1)	     &	-0.4130(1)	&	0.009(1)	\\
        Al(1)	&	4e	&	1	&	-0.2936(2)	  &	-0.2064(2)	     &	-0.4273(1)	&	0.007(1)	\\
        Al(2)	&	2c	&	1	&	0	          &	-0.5	         &	-0.3544(1)	&	0.010(1)	\\
        Al(3)	&	4e	&	1	&	-0.2557(2)	  &	-0.2443(2)	     &	-0.0851(2)	&	0.007(1)	\\
        Al(4)	&	4e	&	1	&	-0.7436(2)	  &	-0.2436(2)	     &	-0.0852(2)	&	0.008(1)	\\
     \hline
\hline
\end{tabular}
  \end{center}
\end{table}


\begin{table}[h]
\caption{Atomic coordinates and equivalent displacement parameters (\AA$^{2}$) for Ce$_{2}$PdAl$_{7}$Ge$_{4}$ determined by single crystal x-ray diffraction at room temperature. U$_{eq}$ is defined as one third of the trace of the orthogonalized U$^{ij}$ tensor.}\label{Table-Pd}
\begin{center}
\begin{tabular}{l@{\hspace{0.7cm}}c@{\hspace{0.7cm}}c@{\hspace{0.7cm}}c@{\hspace{0.7cm}}c@{\hspace{0.7cm}}c@{\hspace{0.7cm}}c}								
		\hline													
        Atom    & Wyck. & Occ.  &	  x	          &	      y	         &	  z	        &   U$_{eq}$    \\
		\hline		
         Pd(1)	&	2a	&	1	&	0	          &	0	             &	0	        &	0.004(1)	\\
         Ce(1)	&	4d	&	1	&	0	          &	0	             &	-0.2611(1)	&	0.005(1)	\\
         Ge(1)	&	2c	&	1	&	0.5	          &	0	             &	-0.2031(1)	&	0.007(1)	\\
         Ge(2)	&	2c	&	1	&	0	          &	0.5	             &	-0.1906(1)	&	0.008(1)	\\
         Ge(3)	&	4e	&	1	&	0.2796(1)	  &	-0.2204(1)	     &	-0.4139(1)	&	0.009(1)	\\
         Al(1)	&	4e	&	1	&	-0.2974(3)	  &	-0.2026(3)	     &	-0.4269(1)	&	0.005(1)	\\
         Al(2)	&	2c	&	1	&	0	          &	-0.5	         &	-0.3572(2)	&	0.007(1)	\\
         Al(3)	&	4e	&	1	&	-0.2562(2)	  &	-0.2438(2)	     &	-0.0890(3)	&	0.007(1)	\\
         Al(4)	&	4e	&	1	&	-0.7430(2)	  &	-0.2430(2)	     &	-0.0895(4)	&	0.008(1)	\\

\hline
\hline
\end{tabular}
  \end{center}
\end{table}

\end{document}